# Optimization of NSLS-II Blade X-ray Beam Position *Monitors: from Photoemission type to Diamond Detector*

**P. Ilinski**

*Submitted to Journal of Physics: Conference Series*

March 2013

**Photon Sciences Directorate**

**Brookhaven National Laboratory**

# Optimization of NSLS-II Blade X-ray Beam Position Monitors: from Photoemission type to Diamond Detector

**P Ilinski**[1]
Brookhaven National Laboratory, Upton NY, USA

E-mail: pilinski@bnl.gov

**Abstract.** Optimization of blade type X-ray Beam Position Monitors (XBPM) was performed for NSLS-II undulator IVU20. Blade material, configuration and operation principle was analyzed to improve XBPM performance. Optimization is based on calculation of the XBPM signal spatial distribution. Along with standard photoemission blades, Diamond Detector Blade (DDB) was analyzed as XBPM signal source. Analyses revealed, that Diamond Detector Blade XBPM would allow overcoming drawbacks of the photoemission type XBPMs.

## 1. Introduction
Photoemission blade X-ray BPMs [1] are standard for most synchrotron radiation facilities. The photoemission XBPMs are non-invasive and can provide high spatial resolution, but they are vulnerable to the background radiation from dipoles and focusing optics due to their high sensitivity to the lower energy photons. Performance of the photoemission type XBPM was analyzed to optimize geometry and configuration of photoemission blades. Optimization is based on calculations of the XBPM signal spatial distribution. An alternative type of a Diamond Detector Blade XBPM [2] was analyzed and compared to the photoemission XBPM.

## 2. Undulator Radiation Source
Optimization of the XBPM was performed for the NSLS-II undulator IVU20. The power density spatial distribution at the minimum undulator gap, corresponding to K=1.8, and at 500 mA storage ring current is shown at figure 1. The power density distribution specifies the XBPM blades high heat load conditions and defines how far the blades should be placed away from the axis of radiation. Blades need to withstand the heat load and should be mechanically stable since blade deformation can be interpreted as the undulator beam motion.

Another important characteristics of undulator radiation for XBPM operation is the spectral flux angular dependence. The flux spectral density of IVU20 at various locations from the axis of undulator radiation at 10m is presented in figure 2 which shows that hard X-rays from high undulator harmonics are radiated at large off-axis angles.

## 3. Photoemission Blade XBPM
Operation of the photoemission XBPM is based on photoemission of electrons from a blade. Tungsten is often used as a blade material for undulator XBPMs due to mechanical properties. The resulting photoemission XBPM signal is convolution of the undulator spectral density and the total electron yield of the blade. Convolution for a portion of the undulator spectra reveals signal distribution for particular undulator harmonic. The signal spatial distributions of tungsten blade for the first and second undulator harmonics of IVU20, K=1.8, at 10m are shown at figures 3 and 4. The signal generated by second undulator harmonic is more intense compare to the signal due to the first harmonic and is localized closer to the axis of undulator radiation. The total signal spatial distribution of tungsten blade XBPM, which includes all undulator harmonics, is presented at figure 5.

[1] To whom any correspondence should be addressed.

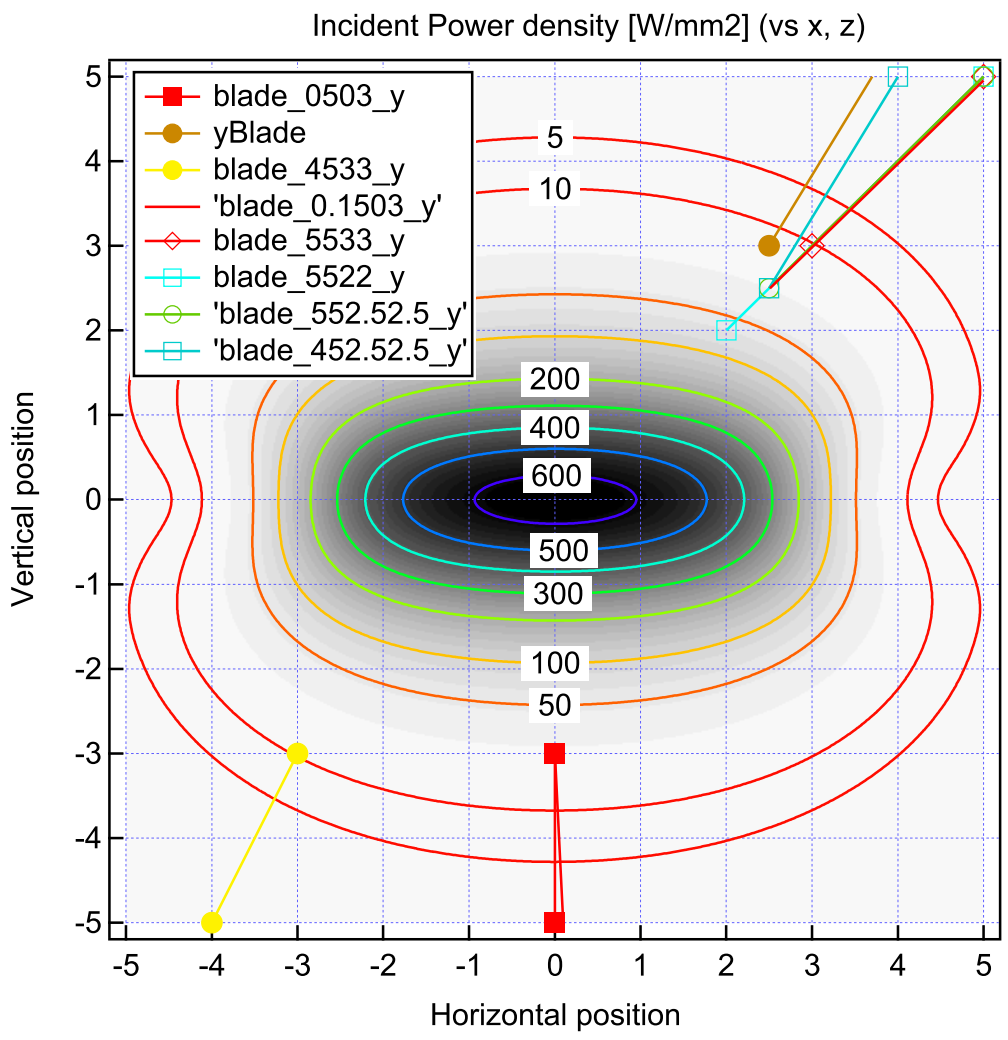

**Figure 1.** Power Density spatial distribution of undulator radiation, IVU20, K=1.8, 10m.

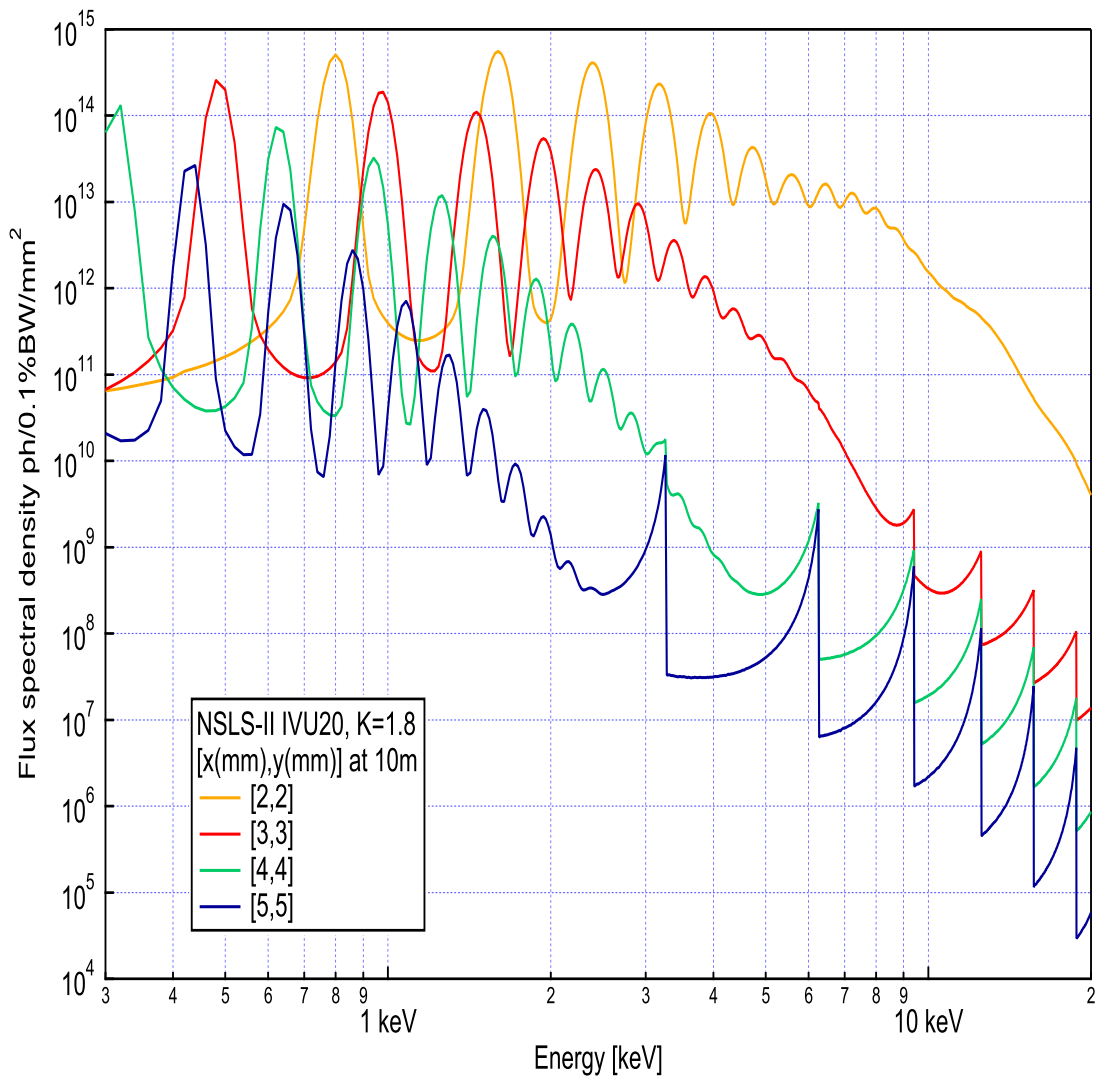

**Figure 2.** Flux Spectral Density angular dependance IVU20, K=1.8, 10m at various [mm, mm] locations from the axis of radiation.

For blades shown at figure 5 the signal level per tungsten blade is reaching few hundreds of micro-amperes at 10m distance for IVU20 at 500 mA current and K=1.8, reducing to tens of micro-amperes for K=0.5. The signal level will increase if blades are extended towards the axis of radiation, but this is limited by the high heat load conditions.

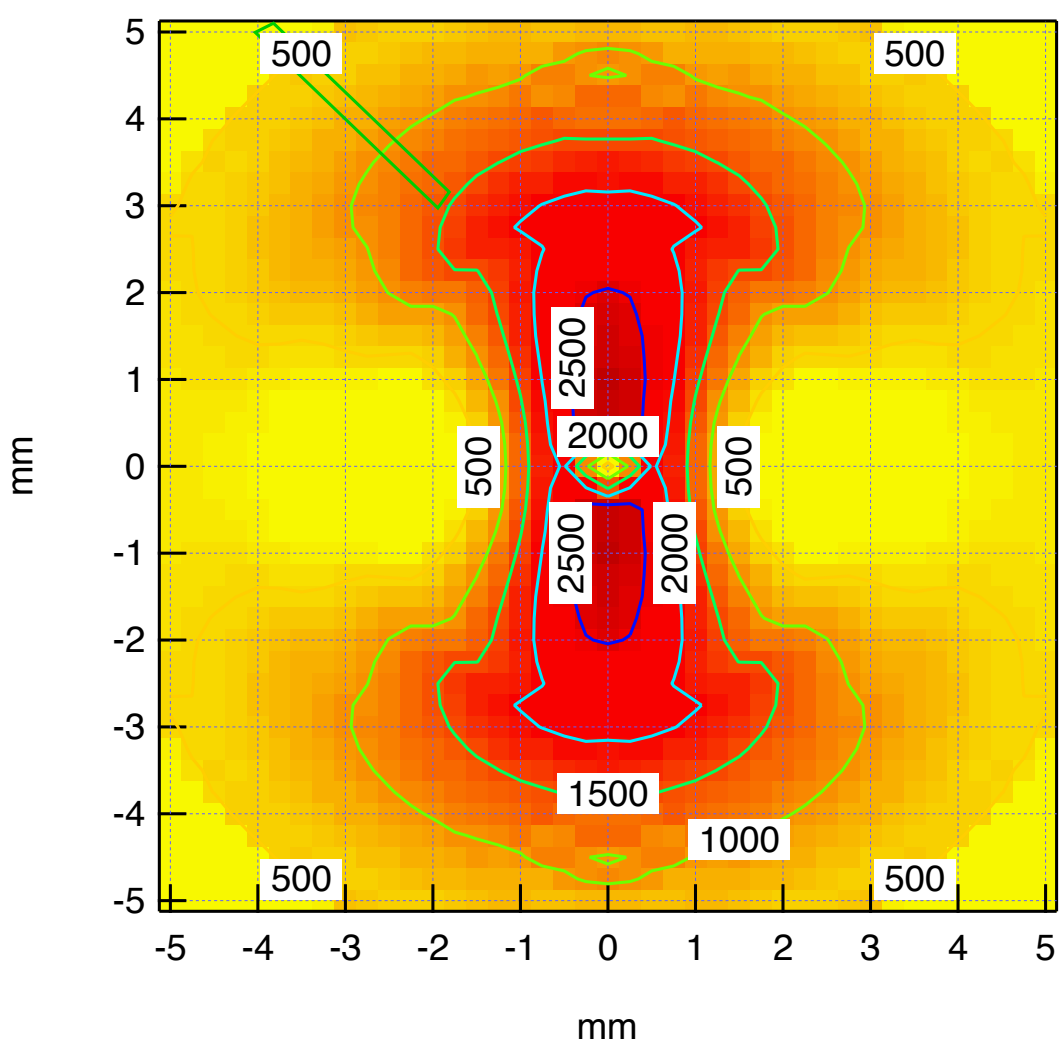

**Figure 3.** The signal spatial distribution for tungsten blade XBPM for the first undulator harmonic, IVU20, K=1.8, 10m.

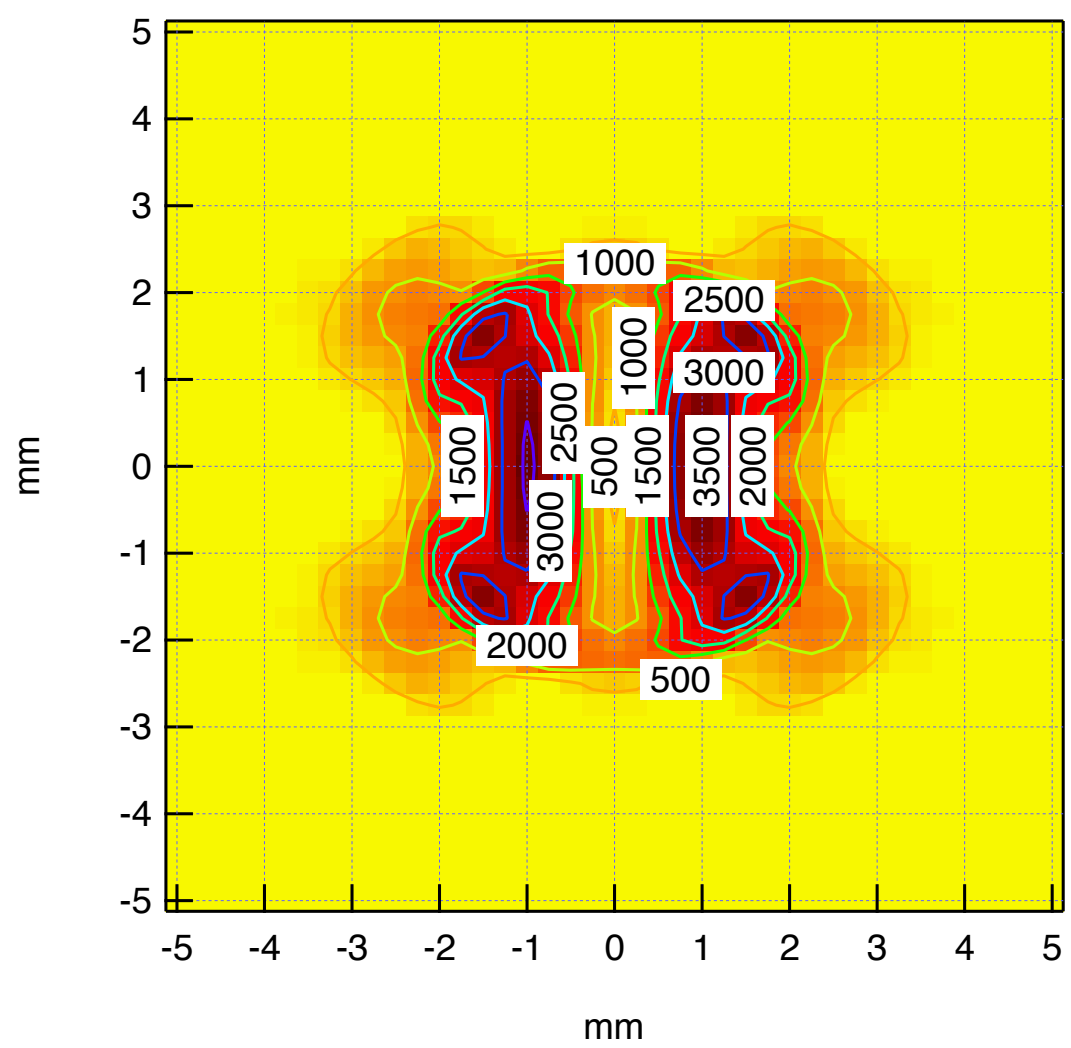

**Figure 4.** The signal spatial distribution for tungsten blade XBPM for second undulator harmonic, IVU20, K=1.8, 10m.

Calibration curves, calculated as (S1-S2)/(S1+S2), where S1 and S2 are signals from the opposite blades, are presented at figure 6 for different configurations of blades indicated in figure 5. As can be seen, the calibration curves do not differ significantly when beam deviation is smaller then 0.5 mm. Therefore, the sensitivity of the photoemission XBPM cannot be improved substantially by optimization of the blade configuration after the blade geometry was defined to satisfy the high heat load conditions.

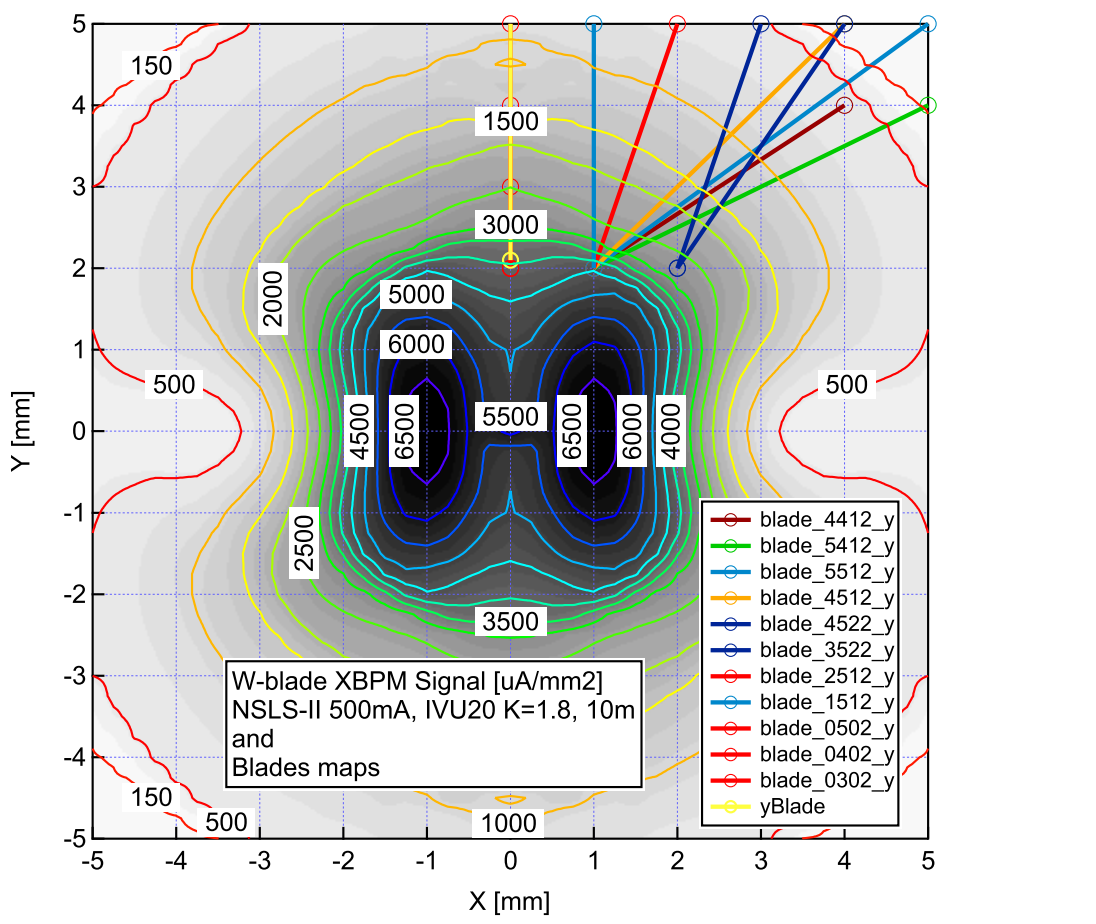

**Figure 5.** Tungsten blade XBPM. The total signal spatial distribution, IVU20, K=1.8, 10m.

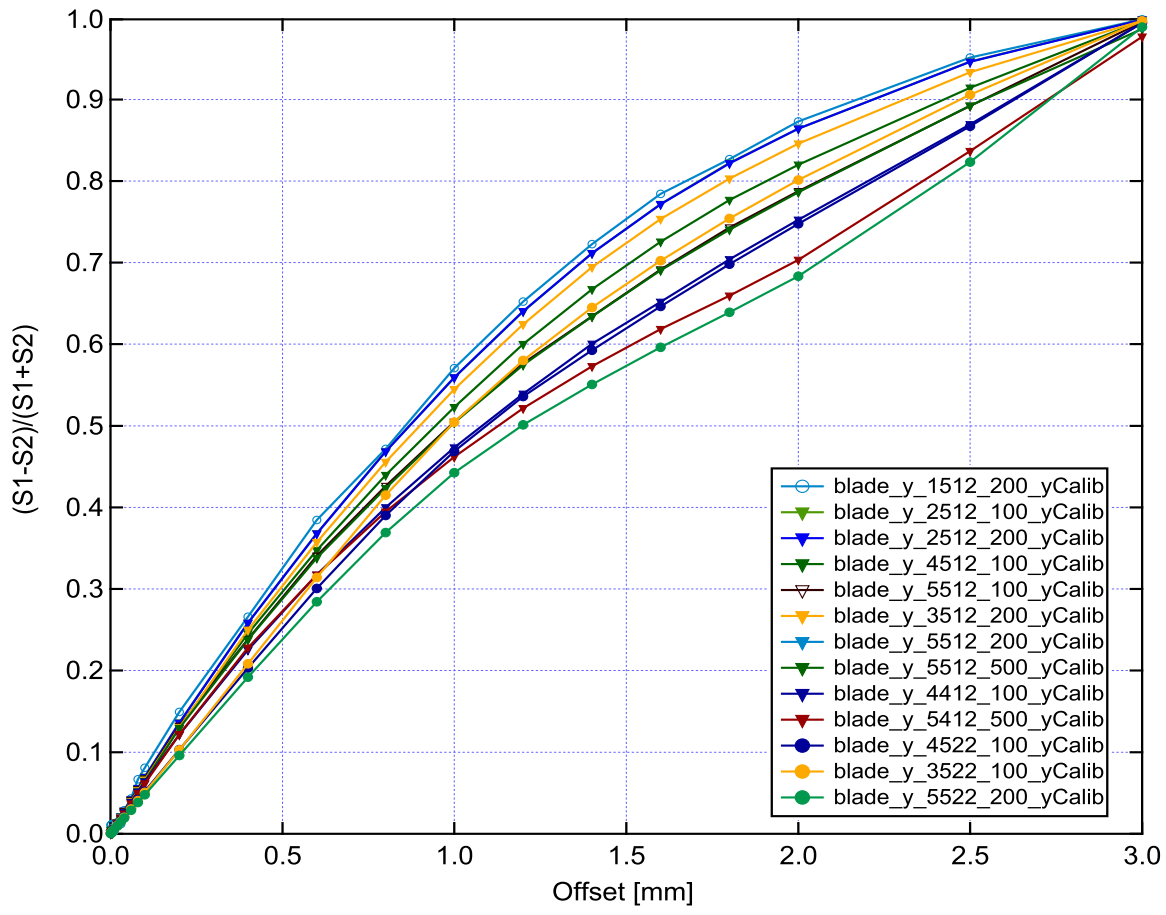

**Figure 6.** Tungsten blade XBPM. The calibration curves for various blades configurations.

## 4. Diamond Detector Blade XBPM

The major deficiencies of the photoemission XBPM are sensitivity to the lower energy photons and dependence on the condition of photoemission surface. Those deficiencies can be overcome by changing the photoemission blade to the Diamond Detector Blade. The DDB XBPM was introduced by H. Aoyagi [2], the layout of the DDB is shown at figure 7. The DDB is positioned along the radiation axis similar to the photoemission blade. Charge carriers generated when photon is absorbed in the diamond detector drift to opposite side electrodes when bias voltage is applied. The signal generated in the diamond detector is proportional to the number and energy of absorbed photons through conversion factor of ~13eV/e-h [3]. This provides an intrinsic discrimination for the lower energy photons. Further discrimination can be achieved by not collecting charge carriers generated by lower energy photons at the front of the DDB by offsetting the electrodes from the edge of the DDB, figure 7. This is similar to an introduction of an X-ray filter in front of the DDB.

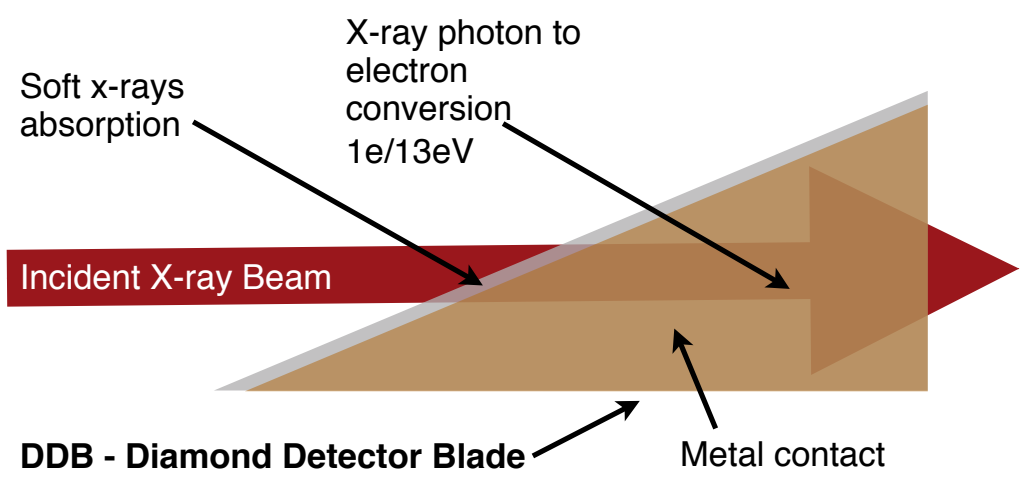

**Figure 7.** Layout of the diamond detector blade of the DDB XBPM.

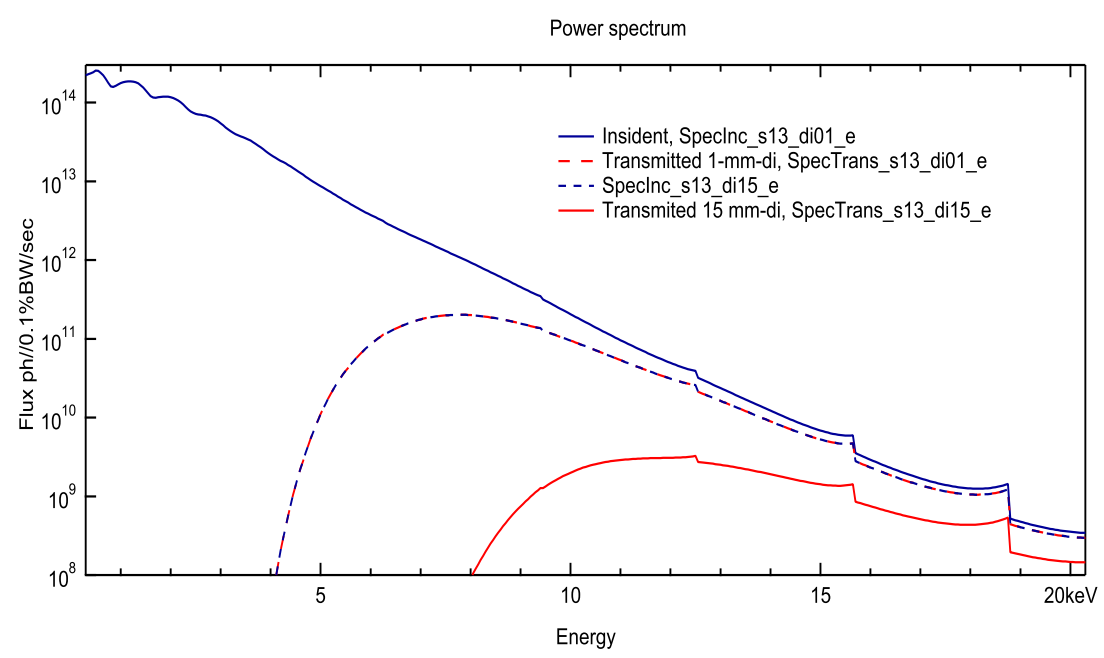

**Figure 8.** Integrated Spectral Flux, IVU20, K=1.8. Incident undulator radiation transmitted after 1mm thick diamond, and transmitted after 16mm thick diamond.

The integrated spectral flux of undulator radiation from IVU20 at K=1.8 is shown at figure 8 along with the integrated spectral flux transmitted after 1mm and 16mm thick diamonds. As can be seen, photons with energies below 4 keV are absorbed in the 1mm thick diamond filter. Varying thickness of the X-ray filter by changing the geometry of side electrodes will make possible to discriminate the lower energy photons background radiation and to control the DDB signal level. Optimization of the side electrodes geometry will also allow control of DDB characteristics such as capacitance.

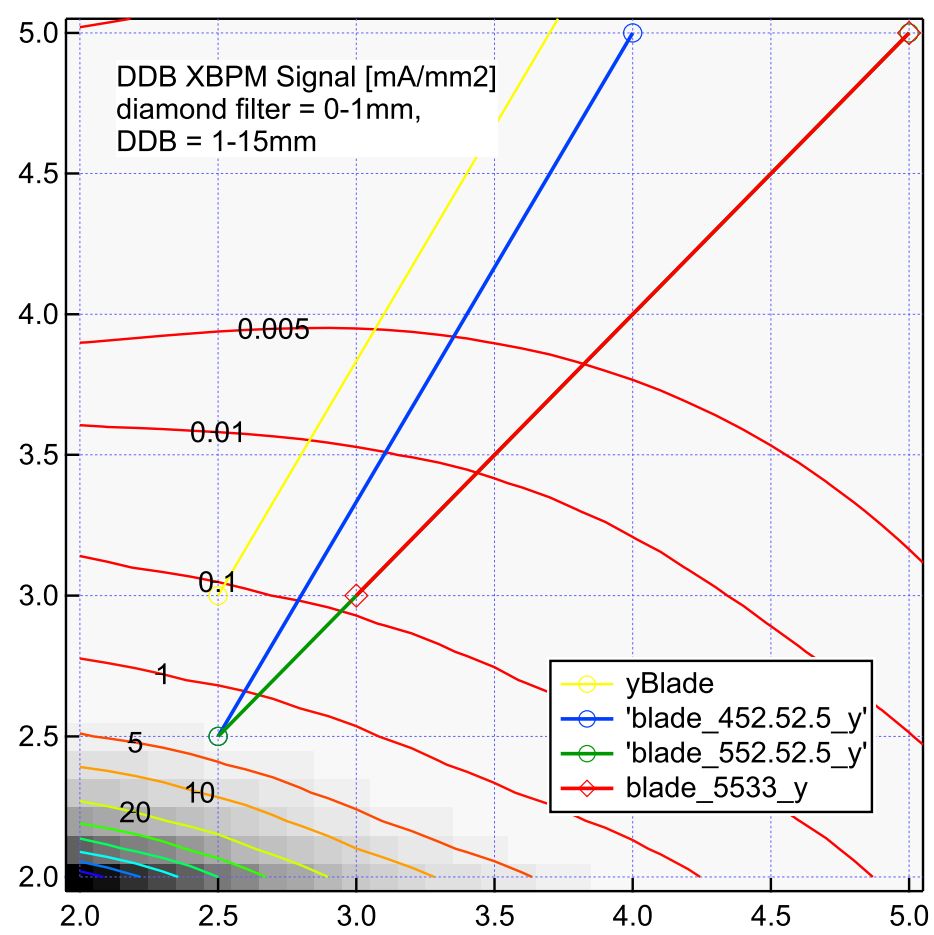

**Figure 9.** The signal spatial distribution of the DDB XBPM, IVU20, K=1.8, 10m, 1mm thick diamond X-ray filter.

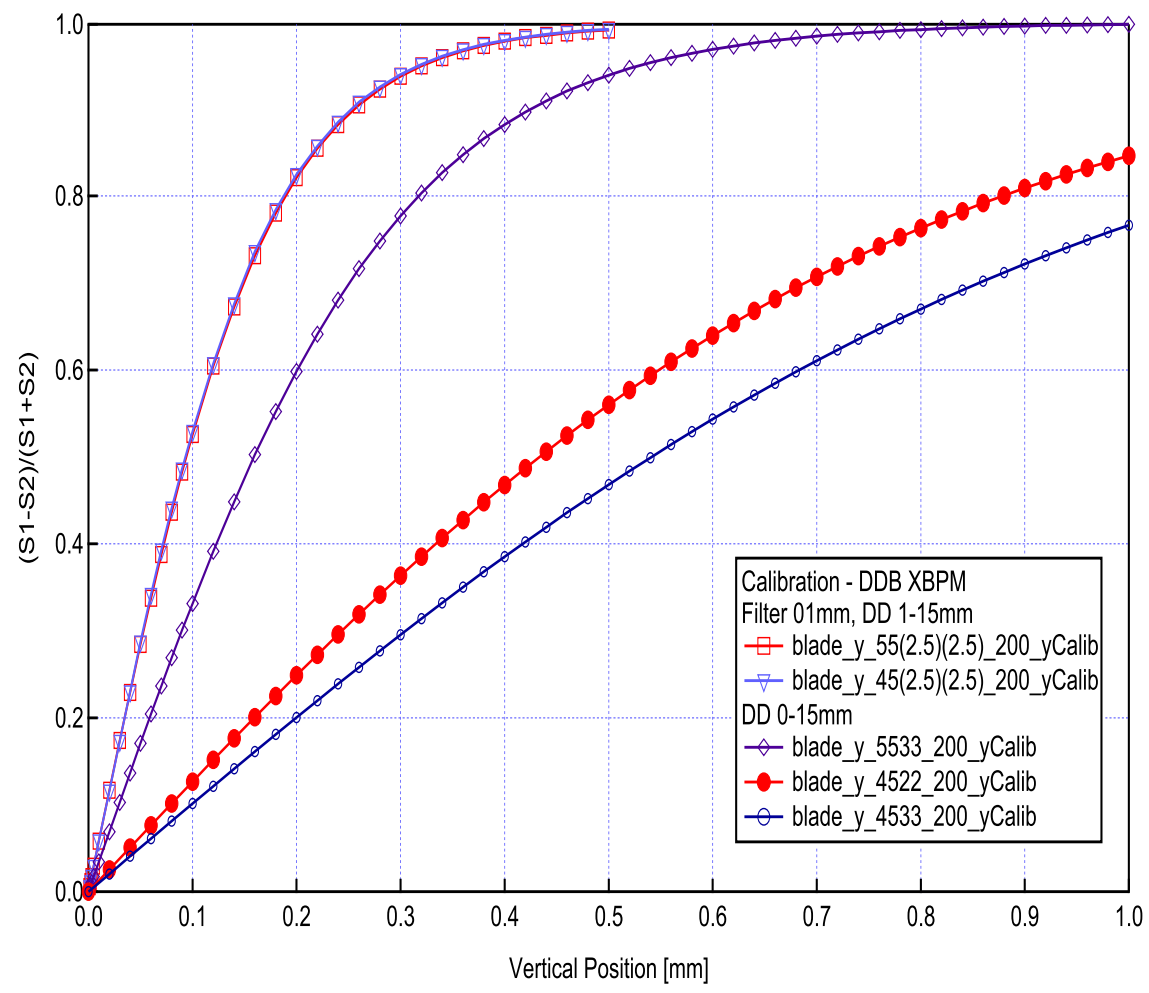

**Figure 10.** The DDB XBPM calibration curves with and without diamond X-ray filter for various blade configurations.

The spatial distribution of the DDB XBPM signal reflects the spatial distribution of the incident radiation power density. The DDB signal is reaching level of few hundreds of milli-amperes for IVU20, K=1.8, at 500 mA storage ring current. When a 1mm thick diamond X-ray filter is introduced in front of the DDB, the signal spatial distribution becomes narrower, figure 9, the signal level drops to hundreds of micro-amperes. The calibration curves of the DDB XBPM with and without diamond X-ray filter are presented at figure 10 for various blade configurations. The sensitivity of the DDB XBPM without X-ray filter is two times higher compare to the tungsten blade XBPM, figure 6. The sensitivity of the DDB XBPM with X-ray filter depends on the blade configuration and is 6-8 times higher compare to the tungsten blade XBPM.

## 5. Conclusion

A noninvasive type of the white beam Diamond Detector Blade XBPM was analyzed and compared to the photoemission XBPM. The choice of the Diamond Detector Blade instead of the photoemission blade as XBPM signal source allows discrimination of the lower energy background photons. The discrimination is achieved due to proportionality of the DDB signal to the energy of absorbed photon, and by additional X-ray filtering through modification of the DDB side electrodes.